\documentclass{aa}  

\usepackage{graphicx}
\usepackage{amsmath,amssymb,latexsym}
\usepackage{natbib}
\usepackage{url}
\usepackage{graphicx}

\begin{document}

\title{On the mass of gas giant planets}

\subtitle{Is Saturn a failed gas giant?}

\author{Ravit Helled}

   \institute{Center for Theoretical Astrophysics \& Cosmology, 
Institute for Computational Science, \\
University of Zurich, Switzerland.\\ }

\abstract
{The formation history of giant planets inside and outside the solar system remains unknown. We suggest that runaway gas accretion is initiated only at a mass of $\sim$ 100 M$_{\oplus}$ and that this  mass corresponds to the transition to a gas giant, a planet that its composition is dominated in hydrogen and helium. Delaying runaway accretion to later times (a few Myr) and higher masses is likely to be a result of an intermediate stage of efficient heavy-element accretion (at a rate of $\sim10^{-5}$ M$_\oplus/yr$) that provides sufficient energy to hinder rapid gas accretion. 
This may imply that Saturn has never reached runaway gas accretion, and that it is a "failed giant planet". The transition to a gas giant planet above Saturn's mass naturally explains the differences between the bulk metallicities and internal structures of Jupiter and Saturn.  
The transition mass to a gas giant planets strongly depends on the exact formation history and birth environment of the planets, which are still not well constrained for our Solar System. In terms of giant exoplanets, delaying runaway gas accretion to planets beyond Saturn's mass can explain the transitions in the mass-radius relations of observed exoplanets and the high metallicity of intermediate-mass  exoplanets. }

\keywords{planets and satellites: formation; planets and satellites: gaseous planets, planets and satellites: composition}

\maketitle

\nolinenumbers
\section{Introduction}
Understanding the origin of the outer planets in the solar system, and giant planet formation in general, is a key objective of planetary science \citep[e.g.,][and references therein]{HelledMorby, Helled2014}. 
The exploration of the giant planets in the solar system by the Juno and Cassini spacecraft provided new and exciting data about the interiors of Jupiter and Saturn, respectively \citep[e.g.,][]{Wahl2017a,2021NatAs...5.1103M,Helled2022,2022A&A...662A..18M}.  The new measurements have challenged giant planet formation theory and have led to the construction of new formation and internal structure models. It is now more accepted that both planets have fuzzy cores, with the core being extended up to 60\% of the planetary radius.
While the exact composition of the planets is not well-determined due to the degenerate nature of structure models and the inference of the composition, Saturn is more metal-rich than Jupiter in relative terms \citep[e.g.,][]{2021PSJ.....2..241N}\footnote{This is in fact can also be seen from the mass-radius relation of the planets \citep[e.g.,][]{2020NatRP...2..562H}}. 
\par

Often, Saturn is treated as being simply a small version of Jupiter in terms of formation and interior structure. However, the two planets significantly differ in their inferred bulk composition and the extension of their cores. In addition, giant planet formation theory predicts that gap formation typically occurs at Jupiter's mass \citep{2017ApJ...835..146D}, which raises the question: what prevented Saturn from growing further in mass? This could be explained if the gas disk disappeared exactly at the right time when Saturn reached its mass, but this is a fine tuning argument which is rather unsatisfying.   
\par 

Traditionally, giant planet formation theory separated the formation of Jupiter and Saturn (the so called "gas giants") from that of Uranus and Neptune (the so called "ice giants"), as it is assumed that Uranus and Neptune grew too slowly to reach the phase of runaway gas accretion, and are therefore "failed giant planets". 
Here we suggest that the transition to hydrogen-helium (H-He) dominated planets, which is associated with undergoing runaway gas accretion during formation, occurs at a mass of $\sim$ 100 Earth masses (M$_{\oplus}$). In that case, Saturn, like Uranus and Neptune, never reached runaway accretion and is also a "failed giant planet". 
The transition to runaway gas accretion at this mass offers interesting explanations for some of the observed features of Jupiter and Saturn and the mass-radius relation of exoplanets.  

\section{Slow giant planet formation and delaying runaway gas accretion} 
In the standard model for giant planet formation, first a heavy-element core is formed followed by gradual solid and gas accretion until  runway gas accretion is reached, leading to a rapid gas accretion \citep{Pollack1996}. 
For a while, giant planet formation models had to accelerate the formation timescale since accretion rates were too low, and the disk's lifetime was assumed to be $\sim$ 3 Myrs \citep[e.g.,][]{2005A&A...434..343A,Movshovitz2010} in order to be consistent with the average estimated disk lifetime at the time \citep{2009AIPC.1158....3M}.  The possibility of rapid core formation via pebble accretion \citep{Lambrechts2014}, and the fact that recent disk observations imply that gaseous disks can have lifetimes beyond 3 Myr \citep[e.g.,][]{2021ApJ...921...72M} relaxes the need for rapid formation of giant planets. It is now known that giant planets are less common than small planets \citep[e.g.,][]{2021ApJS..255...14F} and this could be a result of a relatively slow growth of giant planets in comparison to standard disk lifetimes. 
In addition, more traditional formation models suggest that runaway gas accretion occurs at crossover mass, i.e., when the heavy-element mass is comparable to the H-He mass, which occurs at a planetary mass of $\sim$ 30 M$_{\oplus}$, with the mass strongly dependent on the exact model.

Recent formation models of Jupiter suggest important modifications to the old picture of giant planet formation:
(i) Cores are formed more efficiently via pebble accretion \citep[e.g.,][]{Johansen2017}, (ii) the accretion of solids (heavies) and gas (H-He) leads to a gradual distribution of heavy-elements within the planetary deep interior \citep[e.g.][]{Lambrechts2014,2017ApJ...840L...4H},  (iii) additional enrichment of heavies is required to explain the estimated heavy-element masses in Jupiter and Saturn \citep[e.g.,][]{2010ApJ...720.1161L,Shibata2019,2022ApJ...926L..37S}, and (iv) the formation of Jupiter within a few Myr can explain the existence of two reservoirs of small bodies in the early Solar System \citep{2017PNAS..114.6712K}. 
\par

Indeed \citet{2018NatAs...2..873A} suggested that Jupiter's formation has been characterized by core formation dominated by pebble accretion followed by planetesimal accretion and that Jupiter has reached its final mass at $\sim$ 3 Myr after the beginning of the Solar System. It was later shown more systematically that indeed Jupiter's formation timescale can take a few Myrs if runaway gas accretion is delayed by planetesimal accretion for a large range of initial formation locations and planetesimal sizes \citep{Venturini2020}. 
\par 

Efficient heavy-element accretion, of the order of $10^{-5}$ M$_{\oplus}/yr$,  
provides enough energy to hinder runaway gas accretion since it prolongs the planetary cooling \citep[e.g.,][]{2018NatAs...2..873A,Venturini2020,2023arXiv230412788K}.  Such a formation path includes core formation is dominated by heavy-element accretion (and a very small gas accretion), an intermediate phase of accretion of both solids and a steadily increasing accretion of H-He gas, until the protoplanet is massive enough ($\sim$ 100 M$_{\oplus}$) to initiate rapid accretion of gas (H-He) \citep[e.g.,][]{2022Icar..37814937H}. 
 During this intermediate phase the planetary envelope remains in 
hydrostatic and thermal equilibrium, with the accreted solids providing the entire radiative luminosity. 
In that case the formation timescale of giant planets can be significantly longer (a few Myr) and in addition, the planetary mass associated with the initiation of runaway accretion is significantly higher.  

The required intermediate phase of heavy-element accretion with a rate of the order of $\sim10^{-5}$ M$_{\oplus}/yr$ lasting for a couple Mys, is not only required to delaying gas accretion, but also provides a mechanism for heavy-element enrichment in gas giant planets. 
While the required accretion rate for this model to work  can be estimated, the source of the solid accretion is yet to be determined.  
\par

There are various ways to delay the onset of runaway accretion, and for the sake of argument, hereafter, we simply assume that after core formation further solids can be accreted without specifying the sizes and origin of the solids  i.e., whether the source is pebbles and/or planetesimals (see section 3 for further discussion).  
Finally, also the nature of the so-called runaway accretion phase is not fully understood and is still a topic of intensive investigation.
This phase could be a rapid hydrodynamic collapse of
the envelope or a quasi-hydrostatic equilibrium contraction occurring on a  Kelvin-Helmholtz timescale.

Figure 1 (top panel) shows a sketch of the planetary growth and the expected associated timescales of the different phases.  The bottom panel shows the heavy-element mass fraction in the planet as a function of planetary mass (until a Jupiter mass is reached) in this formation scenario. 
This figure should be taken as a qualitative demonstration of the formation scenario, and includes a shaded region representing the large modeling uncertainty. 
However, {the figure} shows that a longer formation timescale due to an extended second phase can explain the extension of Jupiter's fuzzy core that stops at $\sim$ 100 M$_{\oplus}$ as predicted from structure models. In addition, it  naturally explains why Saturn is has a higher metallicity (heavy-element mass fraction) than Jupiter. 
Note that this mass s significantly higher than the traditional crossover mass of $\sim$ 20-30 M$_{\oplus}$ traditionally assumed in giant planet formation models \citep[e.g.,][and references therein]{Stevenson1982,Bodenheimer1986,Pollack1996,2014ApJ...786...21P,Helled2014}.
\par

A formation scenario in which the initiation of runaway gas accretion takes place at planetary masses of $\sim$ 100 M$_{\oplus}$ implies that Saturn might be a "failed giant planet", i.e., a planet that has never gone through runaway accretion. In this case, no gap opening is required to explain Saturn's final mass, instead, it suggests that Saturn's growth took a few Myrs, and the gas disk disappeared before it could enter the phase of rapid gas accretion that leads to the formation of a gas giant planet.  
It should be noted that the exact planetary mass at which runaway accretion begins depends on the assumed parameters and therefore has a non-negligible uncertainty as indicated by the gray region around Saturn's mass.  The uncertainty is estimated based on different formation models \citep[e.g.,][]{Venturini2020, Valletta2020} but in fact could be larger since the exact composition of the planet and the mass at which runaway accretion is initiated can vary depending on the local formation conditions such as the type of the accreted solids and their distribution within the disk, the planetary  formation location and potential migration, and the disk's physical parameters (temperature, pressure, viscosity). 
\par 

If Saturn never reached the runaway gas accretion phase, it means that  the enrichment of its envelope is a result of heavy-element accretion during the second prolonged phase of accretion and/or mixing of primordial composition gradients that can enrich the outer envelope \citep[e.g.,][]{Vazan2018,Mueller2020}.  
Indeed, Saturn interior models imply that Saturn has an extended fuzzy core \citep[e.g.,][]{2021NatAs...5.1103M,2021PSJ.....2..241N}. 
The delay of gas accretion by a few Myr and the transition to a gas giant planet at a higher mass also naturally explains why Uranus and Neptune are heavy-element dominated in composition  \citep[e.g.][]{2022ApJ...931...21V,2014ApJ...789...69H} - as indicated by Fig.~1, at a planetary mass of $\sim$ 15 M$_{\oplus}$, the H-He mass fraction is about 10\%. 

\begin{figure}
\center
       \includegraphics[angle=0,height=5.3cm]{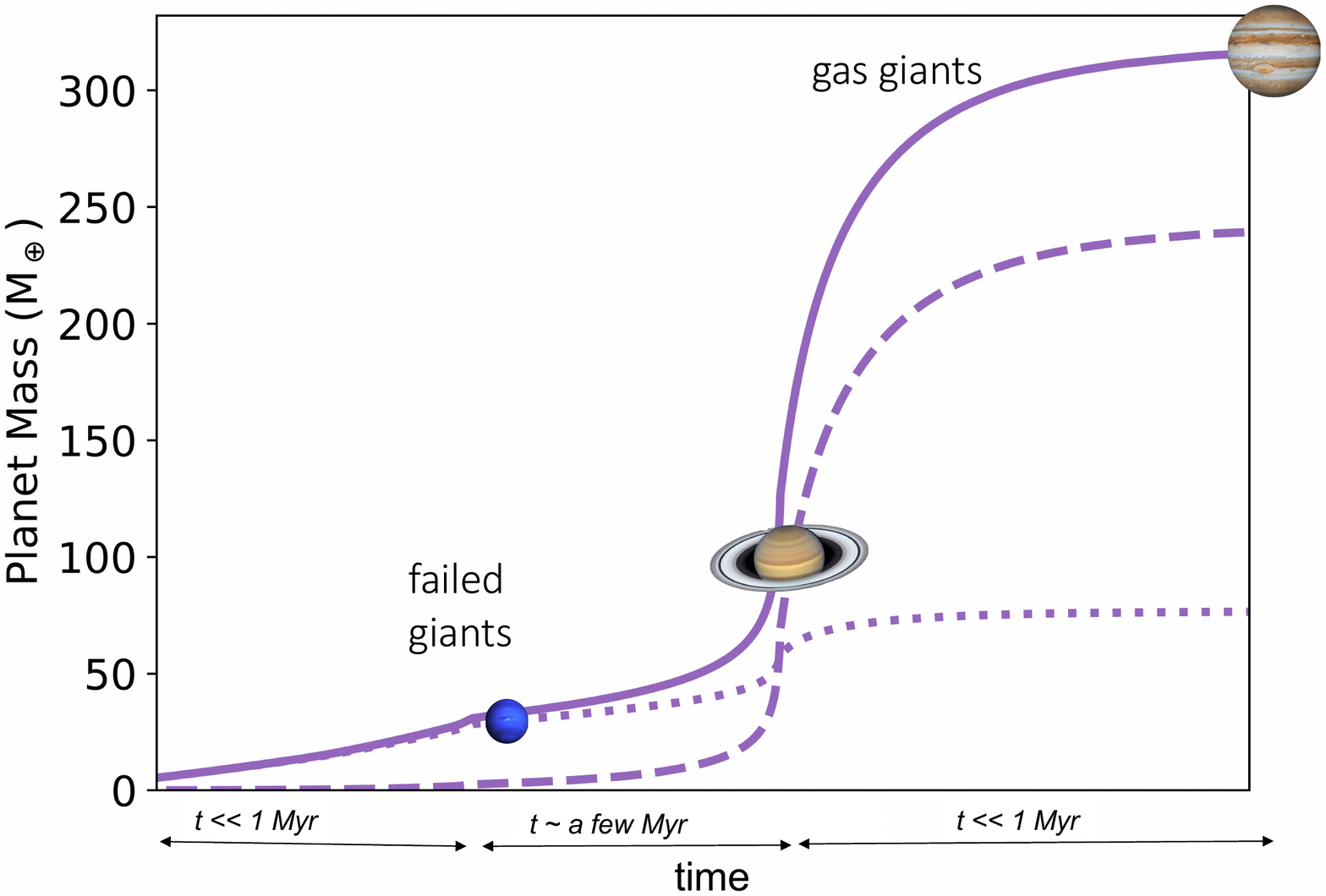}
       \includegraphics[angle=0,height=5.3cm]{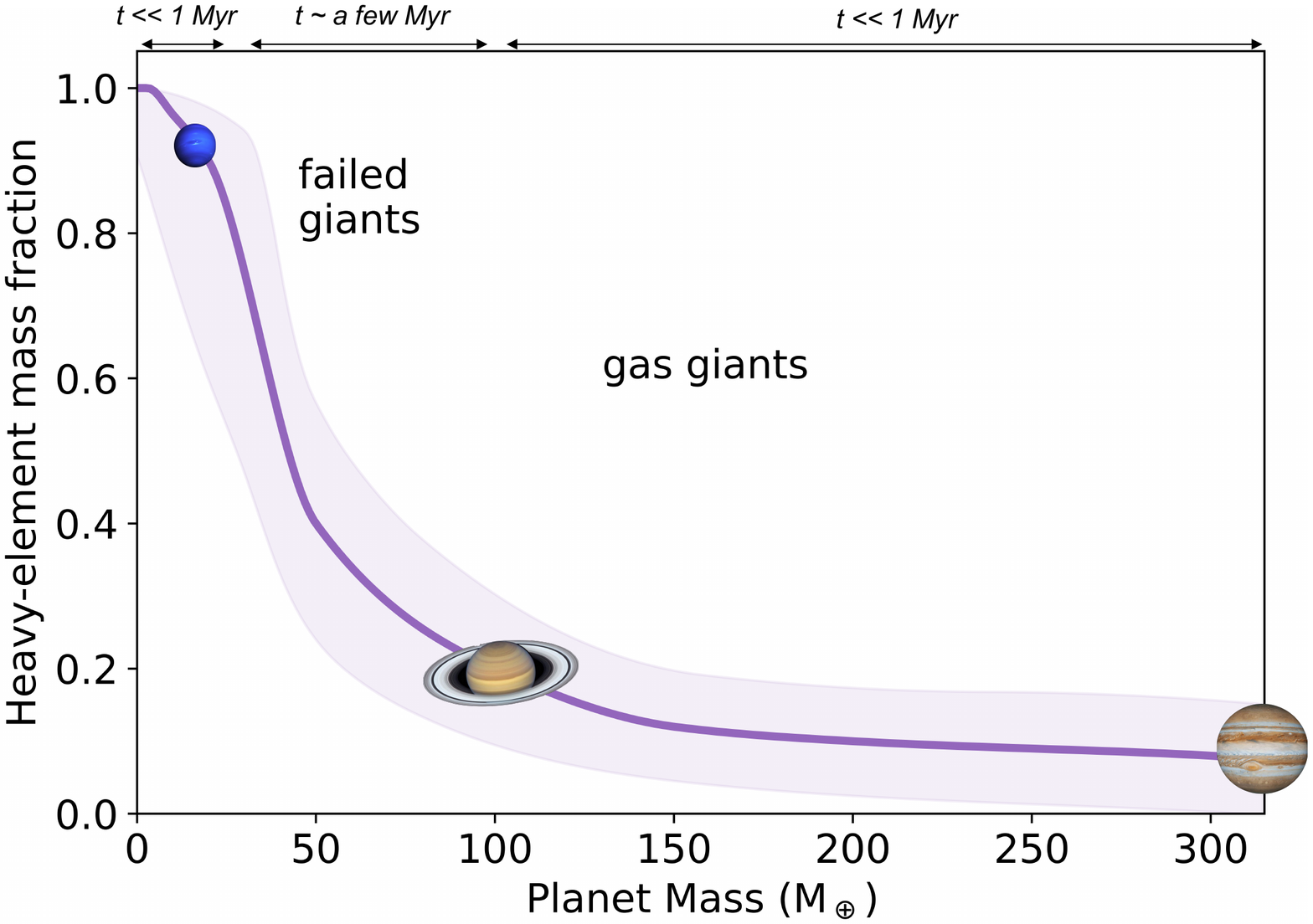}
        \caption{{\bf Top:} Planetary mass as a function of time. The dotted and dashed line correspond to the heavy-element and H- He mass, respectively. 
         The curves is based on formation calculations \citep[see][for details]{2022Icar..37814937H} but includes a large shaded  area to indicate the large uncertainty associated with the details and uncertainties of formation models. 
        The plots clearly show that low-mass planets are expected to be heavy-element dominated in composition while planets more massive than $\sim$ 100 M$_{\oplus}$ are H-He-rich. 
        {\bf Bottom:} The planetary bulk metallicity (given in heavy-element mass fraction) as a function of planetary mass until a Jupiter mass is reached. Interestingly, the expected bulk compositions of the outer planets are consistent with this curve. This model includes planetesimal accretion after core formation which delays cooling and therefore rapid gas accretion \citep{2018NatAs...2..873A,Venturini2020}.  
        Also indicated is the mass where gas accretion is initiated after growth of a few Myr, around a Saturn mass and the order of magnitude of the expected formation timescales.}
            \label{fig:OLS method}
    \end{figure}

\subsection{Giant exoplanets}
From a different point of view, we can use the mass-radius relation (hereafter M-R relation) of exoplanets and their characterization to better understand the nature of giant planets. 
Indeed, in the case of warm Jupiters, which are not highly irradiated, we can estimate the planetary bulk composition \citep[e.g.,][and references therein]{Thorngren2016,2019ApJ...874L..31T,2023A&A...669A..24M}. 
If planets undergo runaway gas accretion, they are expected to be H-He dominated in composition.  
 Interestingly, exoplanet data imply that there is a transition in the M-R relation of exoplanets at a mass of $\sim$ 100 $M_{\oplus}$, which could be associated with the transition into gas giant planets. 
 Beyond this mass, the planets are expected to be H-He dominated in composition, which changes the relation between planetary mass and radius. This is demonstrated in Figure 2 that shows the M-R relation of planets from the PlanetS Catalog{\footnote{\url{https://dace.unige.ch/exoplanets/}} which only includes exoplanets with measurement uncertainties smaller than 25\% in mass and 8\% in radius \citep{2020A&A...634A..43O}. 
 The top panel shows that the M-R relation follows a power-law until the transition mass, around Saturn's mass where the M-R relation becomes flat. Note that the M-R relation of planets that are highly irradiated by their host star ($ F_* > 2 \times 10^8$ erg/s/cm$^{-2}$) is less reliable because their radii are affected by an unknown mechanism and by an unknown magnitude. As a result, these highly irradiated planets are shown in gray. The M-R relation and transition point follows the result of \citet{2017ApJ...834...17C} although similar results an estimated transition mass of 
$\sim$ 120 M$_{\oplus}$ 
 have been inferred by other studies \citep[see e.g.,][]{2013ApJ...768...14W,2015ApJ...810L..25H,2017A&A...604A..83B}\footnote{Using these data, we find that the breakpoint in the M-R relation occurs at M = 120$^{+14}_{-13} M_{\oplus}$}.  
 All these previous statistical analyses identified a transition in the M-R relation of exoplanets occurring at around Saturn mass, and depending on the study, was estimated to occur at a mass of $\sim$95-150 $M_{\oplus}$  (see \citet{2013ApJ...768...14W,2015ApJ...810L..25H,2017ApJ...834...17C,2017A&A...604A..83B} for details).  
\par 

The bottom panel shows the same data with compositional curves of a H-He mixture in a proto-solar ratio for an assumed effective temperature of 1,500 K at age of 4.5 Gyr  \citep{2023A&A...672L...1H} and pure-water with a similar effective temperature \cite{2020A&A...643A.105H}.  One can clearly see that below about Saturn's mass, most of the planets are significantly more compact than predicted from a planet that its composition is dominated by H-He, and that the M-R relation of intermediate-mass exoplanets is more diverse, indicating a range of metallicities (i.e., mass ratios between heavy elements and H-He) that would be a natural outcome of different accretion rates and formation locations. 
\par 

The mass associated with the transition of the M-R relation above Saturn's mass, and the flattening of the M-R relation is caused by the onset of electron degeneracy in hydrogen, and therefore is linked to the planetary bulk composition: the data suggest that beyond this mass the dominating planetary composition is H-He, which is consistent with the initiation of runaway gas accretion. 
Intermediate-mass planets have radii well below the H-He mixture, suggesting that they are composed of other (heavier) elements. In addition, the spread of radii is significantly larger, implying a larger variety in bulk composition. 
\par 

While the conclusion that planets beyond about Saturn's mass are H-He dominated in composition is robust, it is yet to be shown that planets below this mass never reached the stage of runaway gas accreton. 
However, since the expected timescale of runaway gas accretion is rather short ($\ll$ Myr), in order to explain the composition of the intermediate-mass planets, which is not dominated by H-He,  a mechanism to limit gas accretion during runaway would be required (e.g., gas dissipated at the time of gas accretion) or alternatively, one must invoke a post-formation mechanism that results in significant loss of the H-He envelope, such as photo-evaporation, tidal disruption, core-powered mass-loss and giant impacts \cite[e.g.,][]{2012MNRAS.425.2931O, 2018MNRAS.479.5012O,2018MNRAS.476..759G, 2021A&A...648L...1O}. 
\par

 A higher transition mass to giant planets also explains the existence of relatively massive exoplanets (but below Saturn's mass) that are highly enriched with heavy elements \citep[e.g.,][]{2020AJ....160..235B}. 
In addition, delaying runaway gas accretion by a few Myrs indicates that generally giant planets form relatively slowly, in timescales comparable to the disks' lifetime. This makes the formation of giant planets less likely, which is aligned with the relative low occurrence rate of giant exoplanets \citep[e.g.,][]{2015A&A...574A.116R, 2021ApJS..255...14F}.  
It is clear that each individual planet has a unique formation path that depends on the exact formation conditions such as formation location, metallicity, disk lifetime, etc but to first order, a delayed gas accretion offers a simple solution for some key challenges in giant planet theory.

\begin{figure}
\center
       \includegraphics[angle=0,height=6.5cm]{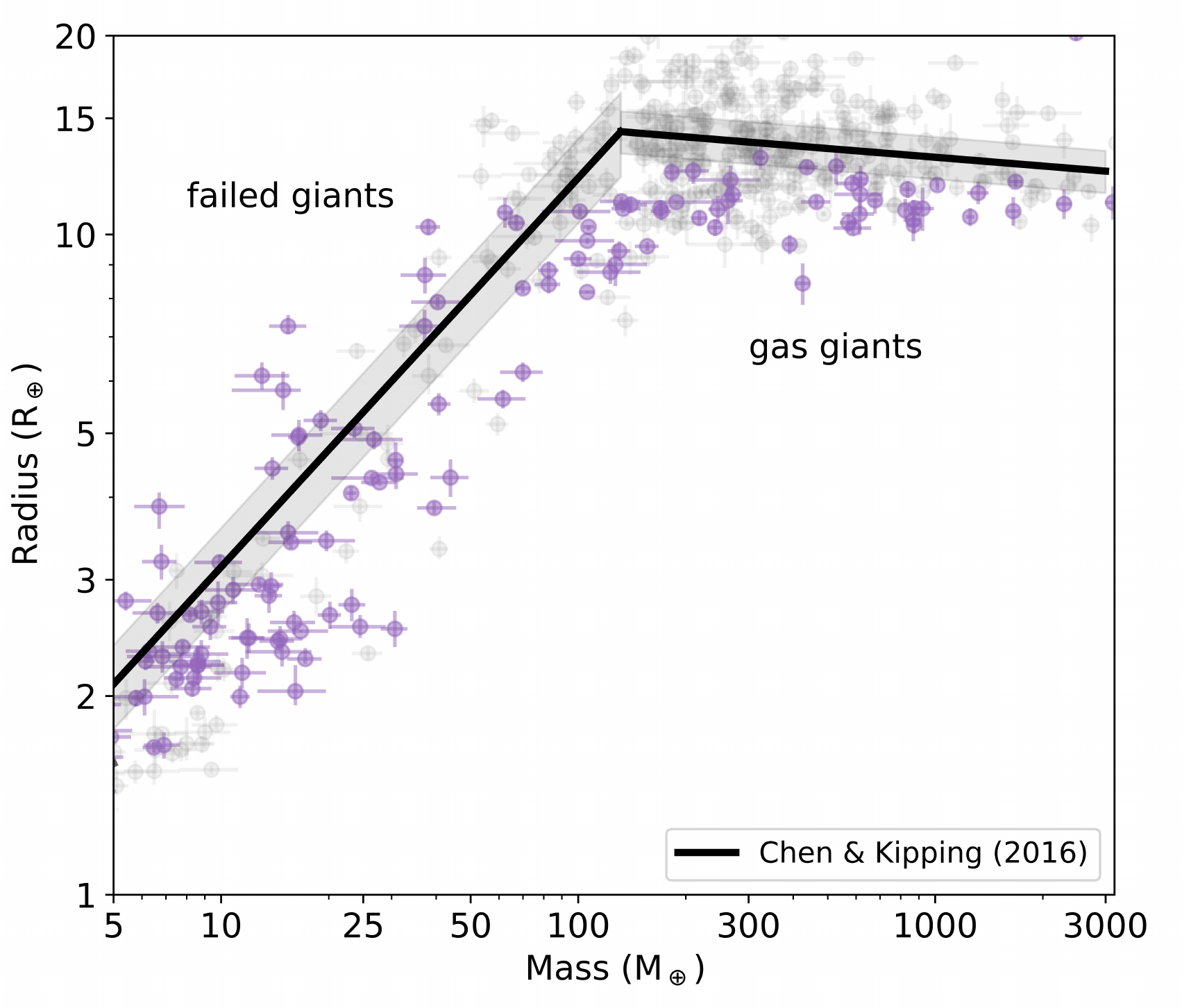}
       \includegraphics[angle=0,height=6.5cm]{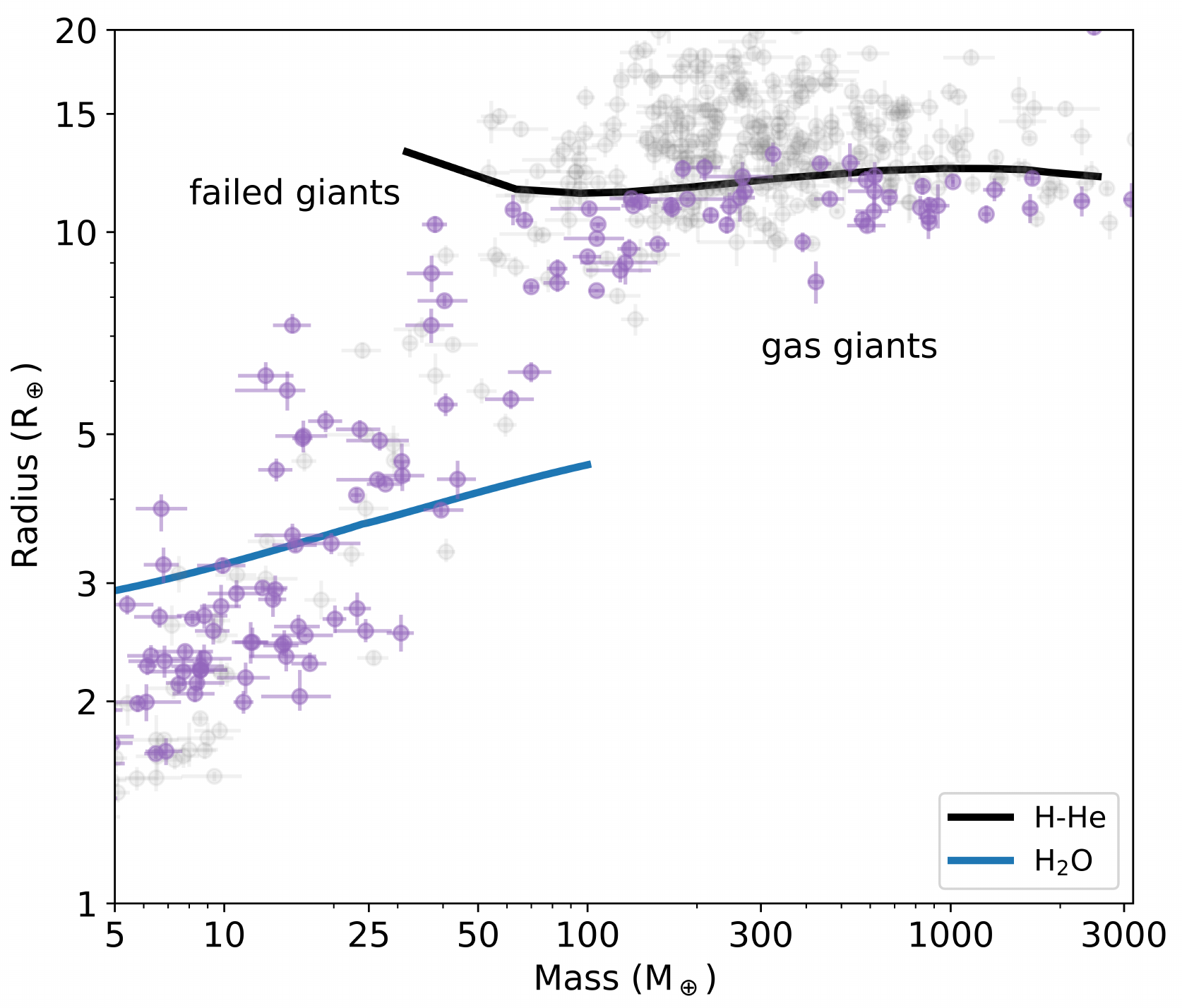}
        \caption{{\small Mass-Radius (M-R) relation of exoplanets from the PlanetS Catalog (\url{https://dace.unige.ch/exoplanets/}). Planets that are highly irradiated by their host star ($ F_* > 2 \times 10^8$ erg/s/cm$^{-2}$) are shown in gray as  their radii are inflated. 
         {\bf Top:} The planets are shown with a M-R relation fit from \citet{2017ApJ...834...17C} 
        and the associated uncertainly in the M-R relation slope. The change in the M-R relation is found to occur at a mass of $\sim$ 120 $M_{\oplus}$. 
        {\bf Bottom:} The planets are shown with composition curves of H-He in a proto-solar ratio (black) and water (blue) assuming an effective temperature of 1,500 K. It is clear that planets below $\sim$Saturn mass are much more compact than planets with a composition that is dominated by H-He.}}
        \label{fig:OLS method}
    \end{figure}


\section{Discussion \& Conclusions}
The picture presented here is clearly oversimplified. 
First, giant planets form at different locations where both the pebble isolation mass and the planetesimal accretions rates are expected to change. 
Therefore the final composition and structure of giant planets would depend on the exact accretion rates as well as the disk's metallicity and lifetime.  It should also be noted that the heavy-element mass in the planet and the extension of the core depends on the local conditions during the formation process. 
In addition, the mutual growth of the planets could affect the accretion rates and the formation timescale.  

Second, the arguments given above are very general and the suggested formation path (and relation between planetary metallicity and mass) was based on a Jupiter formation models. 
In reality, each planet would have a unique path, even if qualitatively similar to the one presented here. For example, since Saturn formed at a larger radial distance than Jupiter, the accretion rates would change accordingly, resulting in a different final heavy-element to H-He ratio as well as a different formation timescale. Since the solid surface density is decreasing with increasing radial distance, smaller accretion rates and therefore longer formation timescale is expected at larger radial distances, and in the case of Uranus and Neptune, their growth could easily take a few Myr, unless they  formation process was dominated by pebble accretion. 
Nevertheless, low H-He masses for  Uranus and Neptune are predicted in both cases of planetesimals and pebble accretion - this is because the pebble isolation mass is increasing with radial distance making runaway gas accretion less likely, and because in the case of planetesimals, the accretion rate is lower due to the lower solid surface density of planetesimals with increasing radial distance. Clearly, detailed investigations of the growth of Uranus and Neptune are still required.  

Third, the source (and likelihood) of solid (heavy-element) accretion at the required rate of  $\sim10^{-5}$ M$_{\oplus}/yr$ is rather uncertain, and more research on this topic is needed. The heavy elements could can come in the form of pebbles, planetesimals and/or both. At the moment, both solely pebble accretion and solely planetesimal accretion have shortcomings, and in reality, probably both type of solids play a role in giant planet formation.  
Pebble accretion is thought to be inefficient  prior to gap opening due to radial pumps \citep[e.g.,][]{2018A&A...612A..30B} and therefore is unlikely to be the dominated mechanism. 
In terms the core formation, pebble accretion is efficient only once a small core has already been formed (above Mars-mass) \citep[e.g.,][]{2010A&A...520A..43O}. 
In addition, recent results from fluid-dynamic simulations suggest that pebble accretion is not effective for cores beyond an Earth-mass due to gas flow around the core \citep{2021ApJ...916..109O}. 
Instead, the collisional growth of pebbles produces planetesimals, which accelerate the core's growth \citep{2021ApJ...922...16K}. As a result, it is likely that also during early stages of core formation the heavy-element accretion is determined from a mixture of pebble and planetesimal accretion.  
\par 

Planetesimals can be accreted during migration in a unperturbed planetesimal disk.  \citet{2022ApJ...926L..37S} showed that efficient planetesimal accretion can occur during migration in a massive planetesimal disk with a heavy-element accretion rate of the order of  $\sim10^{-5}$ M$_{\oplus}/yr$.  
This rate, however, corresponds to a case of an unperturbed planetesimal disk, and it would decreases with increasing inclination excitation. 
At the same time, the growing giant planet is unlikely to increase the planetesimals' inclination before rapid gas \citep[e.g.,][]{1999Icar..139..350T}, instead the embryos around the growing planet are likely to control the inclination of  the planetesimal disk. Nevertheless, even if the inclination is increased by orders of magnitude, the  planetesimal accretion rate is expected to decrease only by a factor of a few \citep{2023MNRAS.519.1713S}. 
However, each of the simulations mentioned correspond to different conditions.  It is therefore clear that a more systematic investigation of the conditions resulting the required  accretion rates to delay runaway gas accretion is desirable, and we hope to address this in future research.   
\par

Fourth, our suggested scenario does not explain low metallicity of intermediate-mass planets as well as planets more massive than Jupiter with very high metallicities. These planets, however, are less common, and could be a result of additional mechanisms that affect the planetary composition and structure such as inflation, atmospheric loss, and giant impacts. 
\par

Finally, the transition mass of $\sim$100 M$_{\oplus}$ to become a giant planet (H-He-dominated in composition) is also not set in stone and is expected to vary depending on the exact formation conditions, and therefore in the case of giant exoplanets, it would depend on the formation environment. 
Nevertheless, this formation path predicts a change in composition for masses beyond Saturn's mass, and that below this mass the planetary metallicity is more diverse can be up to 50\%, consistent with exoplanetary data. 

To summarize, we suggest that the onset of rapid gas accretion occurs at at around Saturn mass, and that this mass defines the transition into gas giant planets. 
If true, this would explain:
\begin{enumerate} 
\item The difference in relative enrichment of the outer planets in the Solar System.
\item The fuzzy cores of Jupiter and Saturn. 
\item The transition of the M-R relation of exoplanets at $\sim$ 100 M$_{\oplus}$ and that above this mass the planetary composition is dominated by H-He.
\item The relative low occurrence rate of giant planets. 
\end{enumerate} 


The origin and internal structure of the outer planets in the solar system are still being investigated. Together with exoplanetary data, we can better characterize giant planets objects, and put our Solar System in perspective. While many questions remain open, this is a golden era in giant planet exploration. The upcoming characterization of many giant planets around other stars from JWST and Ariel, the continuous and future exploration of the outer planets in the Solar System, and the ongoing theoretical efforts will provide new insights on the nature of gas giant planets. 

\subsubsection*{Acknowledgements}
I thank S.~M{\"u}ller for great support. I also thank S.~Shibata, D.~Stevenson, K. Batygin, and C.~Valletta for interesting and useful comments as well as an anonymous referee for valuable corrections and suggestions. Finally, I   acknowledge support from the Swiss National Science Foundation (SNSF) via grant 200020\_188460.


\begin{thebibliography}{56}
\expandafter\ifx\csname natexlab\endcsname\relax\def\natexlab#1{#1}\fi

\bibitem[{{Alibert} {et~al.}(2005){Alibert}, {Mordasini}, {Benz}, \&
  {Winisdoerffer}}]{2005A&A...434..343A}
{Alibert}, Y., {Mordasini}, C., {Benz}, W., \& {Winisdoerffer}, C. 2005, \aap,
  434, 343

\bibitem[{{Alibert} {et~al.}(2018){Alibert}, {Venturini}, {Helled}, {Ataiee},
  {Burn}, {Senecal}, {Benz}, {Mayer}, {Mordasini}, {Quanz}, \&
  {Sch{\"o}nb{\"a}chler}}]{2018NatAs...2..873A}
{Alibert}, Y., {Venturini}, J., {Helled}, R., {et~al.} 2018, Nature Astronomy,
  2, 873

\bibitem[{{Bashi} {et~al.}(2017){Bashi}, {Helled}, {Zucker}, \&
  {Mordasini}}]{2017A&A...604A..83B}
{Bashi}, D., {Helled}, R., {Zucker}, S., \& {Mordasini}, C. 2017, \aap, 604,
  A83

\bibitem[{{Bitsch} {et~al.}(2018){Bitsch}, {Morbidelli}, {Johansen}, {Lega},
  {Lambrechts}, \& {Crida}}]{2018A&A...612A..30B}
{Bitsch}, B., {Morbidelli}, A., {Johansen}, A., {et~al.} 2018, \aap, 612, A30

\bibitem[{Bodenheimer \& Pollack(1986)}]{Bodenheimer1986}
Bodenheimer, P. \& Pollack, J.~B. 1986, Icarus, 67, 391

\bibitem[{{Brahm} {et~al.}(2020){Brahm}, {Nielsen}, {Wittenmyer}, {Wang},
  {Rodriguez}, {Espinoza}, {Jones}, {Jord{\'a}n}, {Henning}, {Hobson},
  {Kossakowski}, {Rojas}, {Sarkis}, {Schlecker}, {Trifonov}, {Shahaf},
  {Ricker}, {Vanderspek}, {Latham}, {Seager}, {Winn}, {Jenkins}, {Addison},
  {Bakos}, {Bhatti}, {Bayliss}, {Berlind}, {Bieryla}, {Bouchy}, {Bowler},
  {Brice{\~n}o}, {Brown}, {Bryant}, {Caldwell}, {Charbonneau}, {Collins},
  {Davis}, {Esquerdo}, {Fulton}, {Guerrero}, {Henze}, {Hogan}, {Horner},
  {Huang}, {Irwin}, {Kane}, {Kielkopf}, {Mann}, {Mazeh}, {McCormac}, {McCully},
  {Mengel}, {Mireles}, {Okumura}, {Plavchan}, {Quinn}, {Rabus}, {Saesen},
  {Schlieder}, {Segransan}, {Shiao}, {Shporer}, {Siverd}, {Stassun}, {Suc},
  {Tan}, {Torres}, {Tinney}, {Udry}, {Vanzi}, {Vezie}, {Vines}, {Vuckovic},
  {Wright}, {Yahalomi}, {Zapata}, {Zhang}, \& {Ziegler}}]{2020AJ....160..235B}
{Brahm}, R., {Nielsen}, L.~D., {Wittenmyer}, R.~A., {et~al.} 2020, \aj, 160,
  235

\bibitem[{{Chen} \& {Kipping}(2017)}]{2017ApJ...834...17C}
{Chen}, J. \& {Kipping}, D. 2017, \apj, 834, 17

\bibitem[{{Dong} \& {Fung}(2017)}]{2017ApJ...835..146D}
{Dong}, R. \& {Fung}, J. 2017, \apj, 835, 146

\bibitem[{{Fulton} {et~al.}(2021){Fulton}, {Rosenthal}, {Hirsch}, {Isaacson},
  {Howard}, {Dedrick}, {Sherstyuk}, {Blunt}, {Petigura}, {Knutson}, {Behmard},
  {Chontos}, {Crepp}, {Crossfield}, {Dalba}, {Fischer}, {Henry}, {Kane},
  {Kosiarek}, {Marcy}, {Rubenzahl}, {Weiss}, \& {Wright}}]{2021ApJS..255...14F}
{Fulton}, B.~J., {Rosenthal}, L.~J., {Hirsch}, L.~A., {et~al.} 2021, \apjs,
  255, 14

\bibitem[{{Ginzburg} {et~al.}(2018){Ginzburg}, {Schlichting}, \&
  {Sari}}]{2018MNRAS.476..759G}
{Ginzburg}, S., {Schlichting}, H.~E., \& {Sari}, R. 2018, \mnras, 476, 759

\bibitem[{{Haldemann} {et~al.}(2020){Haldemann}, {Alibert}, {Mordasini}, \&
  {Benz}}]{2020A&A...643A.105H}
{Haldemann}, J., {Alibert}, Y., {Mordasini}, C., \& {Benz}, W. 2020, \aap, 643,
  A105

\bibitem[{{Hatzes} \& {Rauer}(2015)}]{2015ApJ...810L..25H}
{Hatzes}, A.~P. \& {Rauer}, H. 2015, \apjl, 810, L25

\bibitem[{{Helled} \& {Bodenheimer}(2014)}]{2014ApJ...789...69H}
{Helled}, R. \& {Bodenheimer}, P. 2014, \apj, 789, 69

\bibitem[{{Helled} {et~al.}(2014){Helled}, {Bodenheimer}, {Podolak}, {Boley},
  {Meru}, {Nayakshin}, {Fortney}, {Mayer}, {Alibert}, \& {Boss}}]{Helled2014}
{Helled}, R., {Bodenheimer}, P., {Podolak}, M., {et~al.} 2014, in Protostars
  and Planets VI, ed. H.~{Beuther}, R.~S. {Klessen}, C.~P. {Dullemond}, \&
  T.~{Henning} (University of Arizona Press), 643

\bibitem[{{Helled} {et~al.}(2020){Helled}, {Mazzola}, \&
  {Redmer}}]{2020NatRP...2..562H}
{Helled}, R., {Mazzola}, G., \& {Redmer}, R. 2020, Nature Reviews Physics, 2,
  562

\bibitem[{{Helled} \& {Morbidelli}(2021)}]{HelledMorby}
{Helled}, R. \& {Morbidelli}, A. 2021, in ExoFrontiers; Big Questions in
  Exoplanetary Science, ed. N.~{Madhusudhan}, 12--1

\bibitem[{{Helled} \& {Stevenson}(2017)}]{2017ApJ...840L...4H}
{Helled}, R. \& {Stevenson}, D. 2017, \apjl, 840, L4

\bibitem[{{Helled} {et~al.}(2022{\natexlab{a}}){Helled}, {Stevenson}, {Lunine},
  {Bolton}, {Nettelmann}, {Atreya}, {Guillot}, {Militzer}, {Miguel}, \&
  {Hubbard}}]{2022Icar..37814937H}
{Helled}, R., {Stevenson}, D.~J., {Lunine}, J.~I., {et~al.} 2022{\natexlab{a}},
  \icarus, 378, 114937

\bibitem[{{Helled} {et~al.}(2022{\natexlab{b}}){Helled}, {Werner}, {Dorn},
  {Guillot}, {Ikoma}, {Ito}, {Kama}, {Lichtenberg}, {Miguel}, {Shorttle},
  {Tackley}, {Valencia}, \& {Vazan}}]{Helled2022}
{Helled}, R., {Werner}, S., {Dorn}, C., {et~al.} 2022{\natexlab{b}},
  Experimental Astronomy, 53, 323

\bibitem[{{Howard} \& {Guillot}(2023)}]{2023A&A...672L...1H}
{Howard}, S. \& {Guillot}, T. 2023, \aap, 672, L1

\bibitem[{{Johansen} \& {Lambrechts}(2017)}]{Johansen2017}
{Johansen}, A. \& {Lambrechts}, M. 2017, Annual Review of Earth and Planetary
  Sciences, 45, 359

\bibitem[{{Kessler} \& {Alibert}(2023)}]{2023arXiv230412788K}
{Kessler}, A. \& {Alibert}, Y. 2023, arXiv e-prints, arXiv:2304.12788

\bibitem[{{Kobayashi} \& {Tanaka}(2021)}]{2021ApJ...922...16K}
{Kobayashi}, H. \& {Tanaka}, H. 2021, \apj, 922, 16

\bibitem[{{Kruijer} {et~al.}(2017){Kruijer}, {Burkhardt}, {Budde}, \&
  {Kleine}}]{2017PNAS..114.6712K}
{Kruijer}, T.~S., {Burkhardt}, C., {Budde}, G., \& {Kleine}, T. 2017,
  Proceedings of the National Academy of Science, 114, 6712

\bibitem[{{Lambrechts} \& {Johansen}(2014)}]{Lambrechts2014}
{Lambrechts}, M. \& {Johansen}, A. 2014, \aap, 572, A107

\bibitem[{{Li} {et~al.}(2010){Li}, {Agnor}, \& {Lin}}]{2010ApJ...720.1161L}
{Li}, S.~L., {Agnor}, C.~B., \& {Lin}, D.~N.~C. 2010, \apj, 720, 1161

\bibitem[{{Mamajek}(2009)}]{2009AIPC.1158....3M}
{Mamajek}, E.~E. 2009, in American Institute of Physics Conference Series, Vol.
  1158, Exoplanets and Disks: Their Formation and Diversity, ed. T.~{Usuda},
  M.~{Tamura}, \& M.~{Ishii}, 3--10

\bibitem[{{Mankovich} \& {Fuller}(2021)}]{2021NatAs...5.1103M}
{Mankovich}, C.~R. \& {Fuller}, J. 2021, Nature Astronomy, 5, 1103

\bibitem[{{Michel} {et~al.}(2021){Michel}, {van der Marel}, \&
  {Matthews}}]{2021ApJ...921...72M}
{Michel}, A., {van der Marel}, N., \& {Matthews}, B.~C. 2021, \apj, 921, 72

\bibitem[{{Miguel} {et~al.}(2022){Miguel}, {Bazot}, {Guillot}, {Howard},
  {Galanti}, {Kaspi}, {Hubbard}, {Militzer}, {Helled}, {Atreya}, {Connerney},
  {Durante}, {Kulowski}, {Lunine}, {Stevenson}, \&
  {Bolton}}]{2022A&A...662A..18M}
{Miguel}, Y., {Bazot}, M., {Guillot}, T., {et~al.} 2022, \aap, 662, A18

\bibitem[{{Movshovitz} {et~al.}(2010){Movshovitz}, {Bodenheimer}, {Podolak}, \&
  {Lissauer}}]{Movshovitz2010}
{Movshovitz}, N., {Bodenheimer}, P., {Podolak}, M., \& {Lissauer}, J.~J. 2010,
  \icarus, 209, 616

\bibitem[{{M{\"u}ller} \& {Helled}(2023)}]{2023A&A...669A..24M}
{M{\"u}ller}, S. \& {Helled}, R. 2023, \aap, 669, A24

\bibitem[{{M{\"u}ller} {et~al.}(2020){M{\"u}ller}, {Helled}, \&
  {Cumming}}]{Mueller2020}
{M{\"u}ller}, S., {Helled}, R., \& {Cumming}, A. 2020, \aap, 638, A121

\bibitem[{{Nettelmann} {et~al.}(2021){Nettelmann}, {Movshovitz}, {Ni},
  {Fortney}, {Galanti}, {Kaspi}, {Helled}, {Mankovich}, \&
  {Bolton}}]{2021PSJ.....2..241N}
{Nettelmann}, N., {Movshovitz}, N., {Ni}, D., {et~al.} 2021, Planetary Science
  Journal, 2, 241

\bibitem[{{Ogihara} {et~al.}(2021){Ogihara}, {Hori}, {Kunitomo}, \&
  {Kurosaki}}]{2021A&A...648L...1O}
{Ogihara}, M., {Hori}, Y., {Kunitomo}, M., \& {Kurosaki}, K. 2021, \aap, 648,
  L1

\bibitem[{{Okamura} \& {Kobayashi}(2021)}]{2021ApJ...916..109O}
{Okamura}, T. \& {Kobayashi}, H. 2021, \apj, 916, 109

\bibitem[{{Ormel} \& {Klahr}(2010)}]{2010A&A...520A..43O}
{Ormel}, C.~W. \& {Klahr}, H.~H. 2010, \aap, 520, A43

\bibitem[{{Otegi} {et~al.}(2020){Otegi}, {Bouchy}, \&
  {Helled}}]{2020A&A...634A..43O}
{Otegi}, J.~F., {Bouchy}, F., \& {Helled}, R. 2020, \aap, 634, A43

\bibitem[{{Owen} \& {Jackson}(2012)}]{2012MNRAS.425.2931O}
{Owen}, J.~E. \& {Jackson}, A.~P. 2012, \mnras, 425, 2931

\bibitem[{{Owen} \& {Lai}(2018)}]{2018MNRAS.479.5012O}
{Owen}, J.~E. \& {Lai}, D. 2018, \mnras, 479, 5012

\bibitem[{{Piso} \& {Youdin}(2014)}]{2014ApJ...786...21P}
{Piso}, A.-M.~A. \& {Youdin}, A.~N. 2014, \apj, 786, 21

\bibitem[{{Pollack} {et~al.}(1996){Pollack}, {Hubickyj}, {Bodenheimer},
  {Lissauer}, {Podolak}, \& {Greenzweig}}]{Pollack1996}
{Pollack}, J.~B., {Hubickyj}, O., {Bodenheimer}, P., {et~al.} 1996, \icarus,
  124, 62

\bibitem[{{Reffert} {et~al.}(2015){Reffert}, {Bergmann}, {Quirrenbach},
  {Trifonov}, \& {K{\"u}nstler}}]{2015A&A...574A.116R}
{Reffert}, S., {Bergmann}, C., {Quirrenbach}, A., {Trifonov}, T., \&
  {K{\"u}nstler}, A. 2015, \aap, 574, A116

\bibitem[{{Shibata} \& {Helled}(2022)}]{2022ApJ...926L..37S}
{Shibata}, S. \& {Helled}, R. 2022, \apjl, 926, L37

\bibitem[{{Shibata} {et~al.}(2023){Shibata}, {Helled}, \&
  {Kobayashi}}]{2023MNRAS.519.1713S}
{Shibata}, S., {Helled}, R., \& {Kobayashi}, H. 2023, \mnras, 519, 1713

\bibitem[{{Shibata} \& {Ikoma}(2019)}]{Shibata2019}
{Shibata}, S. \& {Ikoma}, M. 2019, \mnras, 487, 4510

\bibitem[{{Stevenson}(1982)}]{Stevenson1982}
{Stevenson}, D.~J. 1982, Annual Review of Earth and Planetary Sciences, 10, 257

\bibitem[{{Tanaka} \& {Ida}(1999)}]{1999Icar..139..350T}
{Tanaka}, H. \& {Ida}, S. 1999, \icarus, 139, 350

\bibitem[{{Thorngren} \& {Fortney}(2019)}]{2019ApJ...874L..31T}
{Thorngren}, D. \& {Fortney}, J.~J. 2019, \apjl, 874, L31

\bibitem[{{Thorngren} {et~al.}(2016){Thorngren}, {Fortney}, {Murray-Clay}, \&
  {Lopez}}]{Thorngren2016}
{Thorngren}, D.~P., {Fortney}, J.~J., {Murray-Clay}, R.~A., \& {Lopez}, E.~D.
  2016, \apj, 831, 64

\bibitem[{{Valletta} \& {Helled}(2020)}]{Valletta2020}
{Valletta}, C. \& {Helled}, R. 2020, \apj, 900, 133

\bibitem[{{Valletta} \& {Helled}(2022)}]{2022ApJ...931...21V}
{Valletta}, C. \& {Helled}, R. 2022, \apj, 931, 21

\bibitem[{{Vazan} {et~al.}(2018){Vazan}, {Helled}, \& {Guillot}}]{Vazan2018}
{Vazan}, A., {Helled}, R., \& {Guillot}, T. 2018, \aap, 610, L14

\bibitem[{{Venturini} \& {Helled}(2020)}]{Venturini2020}
{Venturini}, J. \& {Helled}, R. 2020, \aap, 634, A31

\bibitem[{Wahl {et~al.}(2017)Wahl, Hubbard, Militzer, Guillot, Miguel,
  Movshovitz, Kaspi, Helled, Reese, Galanti, Levin, Connerney, \&
  Bolton}]{Wahl2017a}
Wahl, S., Hubbard, W.~B., Militzer, B., {et~al.} 2017, Geophys. Res. Lett., 44,
  4649

\bibitem[{{Weiss} {et~al.}(2013){Weiss}, {Marcy}, {Rowe}, {Howard}, {Isaacson},
  {Fortney}, {Miller}, {Demory}, {Fischer}, {Adams}, {Dupree}, {Howell},
  {Kolbl}, {Johnson}, {Horch}, {Everett}, {Fabrycky}, \&
  {Seager}}]{2013ApJ...768...14W}
{Weiss}, L.~M., {Marcy}, G.~W., {Rowe}, J.~F., {et~al.} 2013, \apj, 768, 14

\end{thebibliography}
\end{document}